\documentclass[prl,superscriptaddress,twocolumn,nofootinbib,showpacs,preprintnumbers,floatfix]{revtex4}

\usepackage{mathrsfs}
\usepackage{amsfonts,amssymb,amsmath}
\usepackage{graphicx,epsfig}
\usepackage{multirow}
\usepackage{verbatim}
\usepackage{color}
\usepackage{slashed}

\declareslashed{}{/}{0}{-0.05}{E}
\declareslashed{}{/}{0}{-0.05}{P}
\declareslashed{}{/}{0}{-0.10}{\Widehat{P}}

\def\fsu5{$\cal{F}$-$SU(5)$}

\def\bfsu5{$\boldsymbol{\mathcal{F}}$-$\boldsymbol{SU(5)}$}

\def\m1half{$M_{1/2}$}
\def\m3half{$M_{3/2}$}
\def\m32{$M_{32}$}

\def\fb{${\rm fb}^{-1}$~}

\def\mt2{$M_{T2}$}
\def\x2{$\chi^2$}
\def\2b{$M_{T2}b$}

\def\bs0{$B_S^0 \rightarrow \mu^+ \mu^-$}

\def\bea{\begin{eqnarray}}
\def\eea{\end{eqnarray}}

\begin{document}

\title{The Heavy Gluino in Natural No-Scale $\cal{F}$-$SU$(5)}

\author{Thomas Ford}

\affiliation{Department of Chemistry and Physics, Louisiana State University, Shreveport, Louisiana 71115 USA}

\author{Tianjun Li}

\affiliation{CAS Key Laboratory of Theoretical Physics, Institute of Theoretical Physics, 
Chinese Academy of Sciences, Beijing 100190, P. R. China}

\affiliation{ School of Physical Sciences, University of Chinese Academy of Sciences, 
No.19A Yuquan Road, Beijing 100049, P. R. China}

\author{James A. Maxin}

\affiliation{Department of Chemistry and Physics, Louisiana State University, Shreveport, Louisiana 71115 USA}

\author{Dimitri V. Nanopoulos}

\affiliation{George P. and Cynthia W. Mitchell Institute for Fundamental Physics and Astronomy, Texas A$\&$M University, College Station, TX 77843, USA}

\affiliation{Astroparticle Physics Group, Houston Advanced Research Center (HARC), Mitchell Campus, Woodlands, TX 77381, USA}

\affiliation{Academy of Athens, Division of Natural Sciences, 28 Panepistimiou Avenue, Athens 10679, Greece}

%%%%%%%%%%%%%%%%%%%%%%%%%%%%%%%%%%%%%%%%%%%%%%%%%%%%%%%%%%%%%%%%%%%%%%%%%%%%

\begin{abstract}

In light of recent 80--137~$\rm{fb}^{-1}$ results at the LHC Run 2 establishing a lower gluino mass limit of 2.25~TeV, we revisit the supersymmetric GUT model Flipped $SU$(5) with extra vector-like particles, known as $\cal{F}$-$SU$(5), with vanishing No-Scale Supergravity boundary conditions at the string scale of about $2 \times 10^{17}$~GeV, including the supersymmetry breaking $B_{\mu}$ parameter which is strictly enforced as $B_{\mu} = 0$. Given the proportional dependence of all model scales on a single parameter $M_{1/2}$, No-Scale $\cal{F}$-$SU$(5) was shown to possess no electroweak fine-tuning and thus persists as a $natural$ one-parameter model. In this fresh analysis here, we demand consistency with the measured 125~GeV light Higgs boson mass, though we forgo an upper limit on the lightest neutralino relic density. The resulting phenomenology delivers a gluino mass of $M(\widetilde{g}) \lesssim 7.5$~TeV and a lightest supersymmetric particle (LSP) of $M(\widetilde{\chi}_1^0) \lesssim 1.6$~TeV. In order to dilute the relic density down to the WMAP and Planck measurements, we rely upon a single cosmological master coupling $\lambda_6$.
\end{abstract}

%%%%%%%%%%%%%%%%%%%%%%%%%%%%%%%%%%%%%%%%%%%%%%%%%%%%%%%%%%%%%%%%%%%%%%%%%%%%


\preprint{ACT-05-19} %MI-TH-XXXX}%

\maketitle

%%%%%%%%%%%%%%%%%%%%%%%%%%%%%%%%%%%%%%%%%%%%%%%%%%%%%%%%%%%%%%%%%%%%%%%%%%%%

\section{Introduction}

The ATLAS and CMS Collaborations recently released the 80--137~\fb results at the LHC Run 2 recorded from 2017--18, highlighted by no significant deviation beyond the expected Standard Model background~\cite{Moriond_2019}. The dearth of any positive signal of supersymmetry (SUSY) has elevated the lower bound on the gluino mass to 2.25~TeV with regards to the ATLAS and CMS $\widetilde{g} \to t \bar{t}
+ \widetilde{\chi}_1^0$ simplified model scenarios~\cite{ATLAS-CONF-2018-041,CMS-PAS-SUS-19-005}. The advancement of gluino limits above 2~TeV further strains the SUSY model space, providing impetus for phenomenologists to build SUSY models supporting a heavy gluino, yet remain consistent with the measured light Higgs boson mass of $M_h = 125.1 \pm 0.14$~ GeV~\cite{Aad:2012tfa,Chatrchyan:2012xdj,Tanabashi:2018oca} and the WMAP 9-year~\cite{Hinshaw:2012aka} plus 2018 Planck~\cite{Aghanim:2018eyx} observed relic density on the dark matter content in our universe of $\Omega_{DM} h^2 \simeq 0.12$, in addition to satisfying the world average top quark mass of $M_t = 173.1 \pm 0.9$~GeV~\cite{Tanabashi:2018oca}. In this current age of multi-TeV gluino exclusion limits, the intersection of all these empirically validated quantities becomes ever more increasingly difficult. Despite the aforementioned hurdles, we shall present here an intriguing case for a natural SUSY model with no electroweak fine-tuning that does indeed meet the experimental requirements just noted and can also generate a heavy gluino that would not as yet been produced at the LHC Run 2 in sufficient quantities for detection.

The SUSY Grand Unification Theory (GUT) model Flipped $SU(5)$~\cite{Barr:1981qv,Derendinger:1983aj,Antoniadis:1987dx} with additional vector-like multiplets~\cite{Jiang:2006hf} (referred to as flippons~\cite{Li:2010mi}), better known as \fsu5~\cite{Jiang:2008yf, Jiang:2009za}, was shown to possess no electroweak fine-tuning within the construct of vanishing No-Scale Supergravity boundary conditions at the unification scale~\cite{Leggett:2014hha}. The proportional dependence of all SUSY model parameters as well as the electroweak scale $Z$-boson mass $M_Z$ and top quark mass $M_t$ upon a single unified model variable $M_{1/2}$ established No-Scale \fsu5 as a genuine one-parameter model~\cite{Li:2010ws}. This overall rescaling at the leading order in terms of the mass scale $M_{1/2}$ is analogous to the fixing of the Bohr atomic radius in terms of the physical electron mass and charge, by minimization of the electron energy~\cite{Feynman}. In both these instances, the spectrum scales corresponding to fluctuation in the selected constants, whereas the relative internal structure of the model is left intact. The single parameter nature of No-Scale \fsu5 was shown to lead to the rare and highly desirable attribute of no electroweak fine-tuning and thus a rather natural SUSY GUT model~\cite{Leggett:2014hha}. Moreover, No-Scale \fsu5 was shown to be consistent with every SUSY experiment presently collecting data, without exception (For example, see References~\cite{Li:2011ab,Li:2013naa,Li:2016bww}). The natural characteristics of the model and conformity with all ongoing experimental measurements motivated a keen focus in recent years on available LHC Run 2 data within the context of the model's predicted phenomenology~\cite{Li:2017kcq}. The only question was could the model survive potential null LHC findings and thus an escalated severe constraint on the gluino mass?

An analysis of No-Scale \fsu5 in 2016~\cite{Li:2016bww} showed a curious convergence of the observed central value of the light Higgs boson mass, world average top quark mass, and measured WMAP+Planck relic density upon the gluino mass range then being probed at the LHC Run 2. The union of these three critical experimental values in the model occurred in a narrow gluino mass window of $M(\widetilde{g}) = 2.0 - 2.3$~TeV~\cite{Li:2016bww}, justifying the sharp attention paid to the 2017--18 LHC data and any possible hint of a signal~\cite{Li:2017kcq}. However, with the absence of any statistically significant excess events in the 80--137~\fb collision reports~\cite{Moriond_2019}, the revised ATLAS and CMS lower bound on the gluino mass of $M(\widetilde{g}) \ge 2.25$~TeV~\cite{ATLAS-CONF-2018-041,CMS-PAS-SUS-19-005} now presents tension with the No-Scale \fsu5 phenomenology, with little room for discovery if the convergence of $M_h$, $M_t$, and $\Omega_{\widetilde{\chi}_1^0} h^2$ is to be maintained. The sole gluino decay mode in No-Scale \fsu5 for a gluino heavier than about 1~TeV is the $\widetilde{g} \to \widetilde{t}_1 t \to t \bar{t} + \widetilde{\chi}_1^0$ chain~\cite{Li:2013naa}, warranting our primary focus on the ATLAS and CMS simplified model scenarios leading to the 2.25~TeV exclusion boundary on the gluino. The job of the phenomenologist is to monitor and adjust as fresh experimental data warrants, thus we revisit No-Scale \fsu5 here to explore alternative methods of generating a heavy gluino in excess of 2.3~TeV in light of new LHC Run 2 results. To accomplish this, the convergence of $M_h$, $M_t$, and $\Omega_{\widetilde{\chi}_1^0} h^2$ will require tweaking, though given the direct electroweak precision measurements of $M_h$ and $M_t$, the modification must clearly involve only the relic density $\Omega_{\widetilde{\chi}_1^0} h^2$.

The maximum gluino mass in No-Scale \fsu5 is about 2.3~TeV~\cite{Li:2016bww} if we wish to strictly adhere to the measured relic density of $\Omega_{DM} h^2 \simeq 0.12$~\cite{Hinshaw:2012aka,Aghanim:2018eyx}, therefore, we shall remove our constraint on $\Omega_{\widetilde{\chi}_1^0} h^2$. The one-parameter version of No-Scale \fsu5 we study here achieves $\Omega_{\widetilde{\chi}_1^0} h^2 \simeq 0.12$ via stau-neutralino coannihilation and thus provides a bino lightest supersymmetric particle (LSP). By allowing the difference $\Delta M(\widetilde{\tau}_1^{\pm}, \widetilde{\chi}_1^0)$ between the light stau $\tau_1^{\pm}$ and LSP $\widetilde{\chi}_1^0$ to increase beyond that which is paramount to preserve the small $\Omega_{\widetilde{\chi}_1^0} h^2$, we can raise $M_{1/2}$, and thus all proportional SUSY masses, generating a heavy gluino well in excess of 2.3~TeV. Apart from precipitating a larger relic density, another crucial side effect of a heavier SUSY spectrum involves a larger light stop, and hence coupled through a large top Yukawa term, a larger light Higgs boson mass $M_h$ as well. To counter this adverse development, a smaller top quark mass $M_t$ is needed in order to suppress $M_h$ to near its measured value. We strive to contain $M_t$ within its world average range, so the end result is that a maximum $M_{1/2}$ is established and therefore also a cap on the gluino and LSP mass in the model space. We shall show that this upper boundary on the gluino mass is about $M(\widetilde{g}) \lesssim 7.5$~TeV and $M(\widetilde{\chi}_1^0) \lesssim 1.6$~TeV for the LSP. Yet, we are still faced with the reality of diluting the larger relic density down to its WMAP+Planck measured value. We recognize two feasible approaches to solve the large relic density problem: (1) Realize the dilution through a uniquely elegant mechanism that engages a single cosmological master coupling $\lambda_6$~\cite{Ellis:2019jha}; or (2) Introduce the Peccei-Quinn mechanism~\cite{Peccei:1977hh} to solve the strong CP problem, and consider the axino as the LSP~\cite{Peccei:1977hh,Peccei:1977ur,Kim:1984yn,Bonometto:1993fx,Covi:1999ty,Covi:2001nw,Covi:2004rb,Brandenburg:2004du,Steffen:2008qp,Baer:2008yd,Tamvakis:1982mw,Nieves:1985fq,Rajagopal:1990yx,Goto:1991gq,Chun:1992zk,Chun:1995hc}. In this paper we shall elaborate upon only the master coupling $\lambda_6$ method and its exclusive relevance to the Flipped $SU(5) \times U(1)_X$ GUT model.

In this analysis we initially briefly review the favorable elements of the No-Scale \fsu5 model and also an overview of the master $\lambda_6$ coupling. We then describe our analytical process related to the calculations. Finally, we outline the resulting phenomenology, introducing the heavy gluino and more massive SUSY spectrum that evades the current LHC exclusion limits on the gluino mass.

\section{The \bfsu5 Model}

We merely provide a superficial review here. For a more in-depth background we refer the interested reader to the wealth of literature published on No-Scale \fsu5 (For example, see Refs.~\cite{Li:2010rz,Li:2010ws,Maxin:2011hy,Li:2011ab,Li:2016bww} and references therein). No-Scale \fsu5 benefits from all of the favorable phenomenology of Flipped $SU(5) \times U(1)_X$~\cite{Nanopoulos:2002qk,Barr:1981qv,Derendinger:1983aj,Antoniadis:1987dx}, such as (i) fundamental GUT scale Higgs representations (not adjoints); (ii) natural doublet-triplet splitting; (iii) suppression of dimension-five proton decay; (iv) a two-step see-saw mechanism for neutrino masses; (v) a valuable theoretical catalyst in No-Scale Supergravity~\cite{Cremmer:1983bf,Ellis:1983sf, Ellis:1983ei, Ellis:1984bm, Lahanas:1986uc}; (vi) a SUSY breaking mechanism that secures a vanishing cosmological constant at tree level which promotes the longevity and cosmological flatness of our Universe~\cite{Cremmer:1983bf}; (vii) natural suppression of CP violation and flavor-changing neutral currents; and (viii)  minimization of the loop-corrected scalar potential via dynamic stabilization with a vital contraction in parameterization freedom.

Mass degenerate superpartners of the Standard Model fields have not been observed, so it can be assumed that SUSY is broken near or beyond the TeV scale. This occurs initially in a hidden sector in the minimal Supergravity model, then propagates into the observable sector by gravitational interactions.  The SUSY breaking process is parameterized by the following universal soft SUSY breaking terms: gaugino mass $M_{1/2}$, scalar mass $M_0$, and trilinear coupling $A_0$. The ratio of the low-energy Higgs vacuum expectation values (v.e.v.'s) tan$\beta$ and sign of the Higgs bilinear mass term $\mu$ remain undetermined. Knowledge of the Z-boson mass $M_Z$ along with minimization of the electroweak Higgs potential determine the magnitude of the $\mu$ term and its complimentary bilinear soft term $B_{\mu}$ or tan$\beta$ during electroweak symmetry breaking. The most elementary No-Scale SUGRA scenario fixes $M_0 = A_0 = B_{\mu} = 0$ at unification, with the entire array of SUSY breaking soft terms evolving from unification down to low-energy from the single gaugino mass parameter $M_{1/2}$. In that event, the SUSY particle spectrum will be proportional to the mass parameter $M_{1/2}$ at the leading order. Given the mandate that $B_{\mu} = 0$, the value of tan$\beta$ is then determined in principle. To incorporate this numerically, a consistency check is completed by floating the value of tan$\beta$ and subsequently identifying those points that naturally converge on a vanishing $B_{\mu}$ parameter. Points successful in this endeavor are then confirmed as adhering to the strict No-Scale SUGRA $M_0 = A_0 = B_{\mu} = 0$ condition.

Two sets of vector-like multiplets derived from local F-Theory model building~\cite{Jiang:2008yf, Jiang:2009za}, dubbed flippons~\cite{Li:2010mi}, are introduced at the TeV scale to aid in realization of string-scale gauge coupling unification~\cite{Jiang:2006hf, Jiang:2008yf, Jiang:2009za}. Revisions to the one-loop gauge $\beta$-function coefficients $b_i$ to account for contributions from the vector-like multiplets split the final $SU(5) \times U(1)_X$ unification at around $2 \times 10^{17}$~GeV (referred to as the $M_{\cal F}$ scale) and secondary $SU(3)_C \times SU(2)_L$ unification near $10^{16}$~GeV (referred to as the $M_{32}$ scale), flattening the $SU(3)$ Renormalization Group Equation (RGE) running ($b_3 = 0$)~\cite{Li:2010ws} from low-energy to the secondary unification $M_{32}$. The $M_2$ and $M_3$ gaugino mass terms are unified into a single mass term $M_5 = M_2 = M_3$~\cite{Li:2010rz}, and hence $\alpha_5 = \alpha_2 = \alpha_3$, at $M_{32}$. The $M_1$ gaugino mass term experiences a small shift due to $U(1)_X$ flux effects~\cite{Li:2010rz} between $M_{32}$ and $M_{\cal F}$~\cite{Li:2010ws}, therefore, $M_1$ is referred to as $M_{1X}$ above $M_{32}$. The scale $M_{\cal F}$ is defined by unification of the couplings $\alpha_5 = \alpha_{1X}$, with unification elevated to near the string scale and Planck mass. The resultant flattening of the $M_3$ dynamic evolution from $M_{32}$ down to the electroweak scale spawns a uniform SUSY mass spectrum ordering of $M(\widetilde{t}_1) < M(\widetilde{g}) < M(\widetilde{q})$, where the light stop and gluino are lighter than all other squarks~\cite{Li:2011ab}. All the vector-like particles are decoupled at low-energy at a common mass scale $M_V$, which is left as a free parameter. Present LHC constraints on vector-like $T$ and $B$ quarks~\cite{atlas-vectorlike} fix the lower limit near 855~GeV, so we situate our absolute lower $M_V$ limit at $M_V \ge 855$~GeV, though we shall only consider much larger $M_V$ scales in this work.

\section{The Master $\lambda_6$ Coupling}

A remarkable consequence of the No-Scale Flipped $SU(5)$ framework is its natural application to the early universe, providing a Starobinsky-like inflationary model~\cite{Ellis:2013xoa} which is consistent with all present cosmological data. Furthermore, the inflaton is identified with one of the endemic four singlets~\cite{Ellis:2017jcp}\cite{Ellis:2018moe}\cite{Ellis:2019jha} that are necessary to provide light neutrino masses and mixings through the double seesaw mechanism~\cite{Georgi:1979dq}. The $\lambda_6 F \bar{H} \phi$ term ($F = \underline{10}$ of matter fields, $\bar{H} = \overline{10}$ of the GUT Higgs multiplet, and $\phi$ the singlet field) generates a mixture of the inflaton with the heavy right-handed sneutrino field $\widetilde{\nu}_R$, as $\nu_R \in F$. In this framework, the inflaton decays, hence reheating, thus gravitino production. Therefore, LSP production through gravitino decays depends explicitly on the $\lambda_6$ coupling, which also determines the light neutrino masses/mixing~\cite{Ellis:2018moe}\cite{Ellis:2019jha}\cite{EGNNO-2019}. The baryon asymmetry of the universe created through the sphaleron-recycling of the lepton asymmetry initiated by the out of thermal equilibrium $\nu_R$ decays also depends directly on the $\lambda_6$ coupling, thus it has been called the master $\lambda_6$ coupling, or the $\lambda_6$ universe~\cite{Ellis:2019jha}. In this $\lambda_6$ universe, the GUT \fsu5 symmetry is unbroken during inflation, getting broken much later. The flaton $\Phi$, $i.e.$ the combination of the GUT-Higgs Standard Model singlets $\nu_H^c$ and $\nu_{\bar{H}}^c$ that are contained in the $H(10)$ and $\bar{H}(\overline{10})$ that acquires a v.e.v. $V$ ($\sim 10^{16}$~GeV), decouples from the thermal bath, and when $T \le m_{\Phi}$ ($\sim 10^4$~GeV) it becomes non-relativistic and eventually dominates the energy density of the universe until it decays. The decay of the flaton generates a $second$ period of reheating where the amount of entropy released by the flaton decay $\Delta$ may be as big as ${\cal O} (10^4)$. As a result, such kind of dilution factor $\Delta$ may allow us to overproduce thermally produced LSPs that eventually may be diluted down to the right $\Omega_{DM} h^2$, and hence open up the SUSY parameter space~\cite{Ellis:2019jha,EGNNO-2019}.

\section{Analytical Procedure}

The unified gaugino mass parameter $M_{1/2}$ is varied from $1200 \le M_{1/2} \le 5500$~GeV, where $M_{1/2} \simeq 1200$~GeV roughly correlates to $M_{\widetilde{g}} \simeq 1600$~GeV, the minimum gluino mass of interest. We consider a large range on the vector-like mass decoupling scale $M_V$, expanding our scan in this analysis to $10 \le M_V \le 1500$~TeV. The ratio of the v.e.v's tan$\beta$ is permitted to span most of its functional range at $2 \le {\rm tan}\beta \le 50$. For the top quark mass $M_t$, we float the value well outside its known world average limits~\cite{Tanabashi:2018oca} to ascertain intrinsic consistency of the top mass with respect to $M_{1/2}$ and thus the SUSY spectrum. The effective implementation compelled extended scanning limits of $169 \le M_t \le 178$~GeV. The results of $g_{\mu} - 2$ for the muon have suggested a positive value for the SUSY breaking term $\mu$, therefore we only consider $\mu > 0$. The null $B_{\mu} = 0$ No-Scale SUGRA theoretical condition at the $M_{\cal F}$ scale is strictly applied as $| B_{\mu} | \le$ 1~GeV, compatible with the induced variation from fluctuation of the strong coupling within the perimeter of its uncertainty, and similarly with the anticipated scale of radiative electroweak corrections.

\begin{figure}[t]
       \centering
        \includegraphics[width=0.5\textwidth]{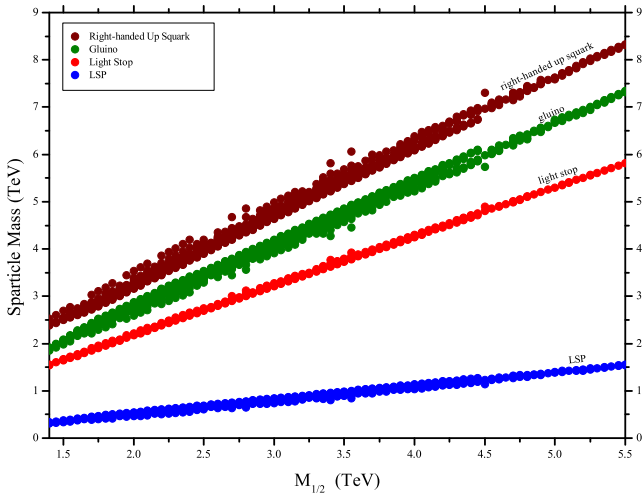}
        \caption{Depiction of SUSY spectrum masses for the LSP neutralino $\widetilde{\chi}_1^0$, light stop $\widetilde{t}_1$, gluino $\widetilde{g}$, and right-handed up squark $\widetilde{u}_R$, as a function of the gaugino mass parameter $M_{1/2}$. All points included adhere to our experimental constraints on the light Higgs boson mass of $124 \le M_h \le 127$~GeV and world average top quark mass of $M_t = 173.1 \pm 0.9$~GeV. No constraint on the LSP relic density $\Omega_{\widetilde{\chi}_1^0} h^2$ is applied here. All points shown here satisfy the No-Scale SUGRA theoretical constraint $|B_{\mu}| \le 1$~GeV.}
        \label{fig:m12}
\end{figure}

\begin{figure}[t]
       \centering
        \includegraphics[width=0.5\textwidth]{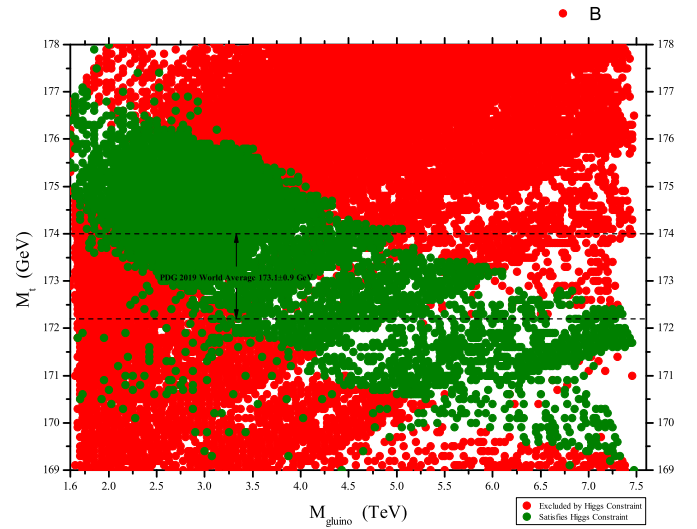}
        \caption{Gluino mass as a function of the top quark mass, distinguished by whether the point satisfies the light Higgs boson mass constraint of $124 \le M_h \le 127$~GeV. Also delineated by the dashed lines is the world average top quark mass of $M_t = 173.1 \pm 0.9$~GeV. No constraint on the LSP relic density $\Omega_{\widetilde{\chi}_1^0} h^2$ is applied here. All points shown here satisfy the No-Scale SUGRA theoretical constraint $|B_{\mu}| \le 1$~GeV. Notice that the viable points consistent with the light Higgs boson mass constraint exit conformity with the world average top quark mass range at about $M_{\widetilde{g}} \simeq 7.5$~TeV.}
        \label{fig:mt}
\end{figure}

\begin{table*}[htp]
  \centering
  \scriptsize
  \caption{The single SUSY breaking soft term $M_{1/2}$, vector-like flippon decoupling scale $M_V$, low energy ratio of Higgs vacuum expectation values (v.e.v.'s) tan$\beta$, top quark mass $M_t$, and relevant SUSY masses for benchmarks points representative of No-Scale \fsu5. Further provided are the two-stage unification scales $M_{32}$ and $M_{\cal F}$. All masses are in GeV. The relic density $\Omega_{\widetilde{\chi}_1^0} h^2$ consists of the calculated abundance for only the LSP $\widetilde{\chi}_1^0$, \underline{before} dilution by the cosmological master coupling $\lambda_6$. Also itemized are the values and branching ratios for the rare-decay processes.}
\label{tab:spectra}
\begin{tabular}{|c|c|c|c||c|c|c|c|c|c||c|c||c|c|c|c|} \hline
$M_{1/2}$ & $M_V$  &  ${\rm tan}\beta$  &  $M_t$  &  $M_{\widetilde{\chi}_1^0}$  &  $M_{\widetilde{\tau}_1^{\pm}}$  &  $M_{\widetilde{t}_1}$ & ${ \bf M_{\widetilde{g}}}$ & $M_{\widetilde{u}_R}$ & $M_h$  & $M_{32}$ & $M_{\cal F}$ & $\Omega_{\widetilde{\chi}_1^0} h^2$ & $\Delta a_{\mu} $ & $ Br(b \to s \gamma)$ & $ Br(B_s^0 \to \mu^+ \mu^-)$ \\ \hline
$	1750	$&$	190,000	$&$	24	$&$	173.8	$&$	440	$&$	461	$&$	1919	$&$	{\bf 2416}	$&$	2879	$&$	124.64	$&$	1.1 \times 10^{16}	$&$	2.4\times 10^{17}	$&$	0.74	$&$	1.7\times 10^{-10}	$&$	3.54\times 10^{-4}	$&$	3.2\times 10^{-9}	$	\\ 	\hline
$	1950	$&$	10,000	$&$	9	$&$	173.1	$&$	447	$&$	682	$&$	2170	$&$	{\bf 2532}	$&$	3445	$&$	125.11	$&$	8.4\times 10^{15}	$&$	3.4\times 10^{17}	$&$	4.17	$&$	4.5\times 10^{-11}	$&$	3.61\times 10^{-4}	$&$	3.1\times 10^{-9}	$	\\ 	\hline
$	1850	$&$	610,000	$&$	23	$&$	173.7	$&$	485	$&$	522	$&$	2025	$&$	{\bf 2606}	$&$	2970	$&$	124.40	$&$	1.1\times 10^{16}	$&$	2.1\times 10^{17}	$&$	1.57	$&$	1.5\times 10^{-10}	$&$	3.56\times 10^{-4}	$&$	3.2\times 10^{-9}	$	\\ 	\hline
$	1950	$&$	1,150,000	$&$	23	$&$	173.6	$&$	524	$&$	556	$&$	2132	$&$	{\bf 2773}	$&$	3093	$&$	124.41	$&$	1.1\times 10^{16}	$&$	1.9\times 10^{17}	$&$	1.36	$&$	1.4\times 10^{-10}	$&$	3.56\times 10^{-4}	$&$	3.2\times 10^{-9}	$	\\ 	\hline
$	2250	$&$	40,000	$&$	10	$&$	173.7	$&$	548	$&$	782	$&$	2458	$&$	{\bf 2978}	$&$	3792	$&$	125.68	$&$	8.5\times 10^{15}	$&$	2.7\times 10^{17}	$&$	5.33	$&$	4.0\times 10^{-11}	$&$	3.61\times 10^{-4}	$&$	3.1\times 10^{-9}	$	\\ 	\hline
$	2200	$&$	160,000	$&$	10	$&$	172.6	$&$	563	$&$	767	$&$	2395	$&$	{\bf 2999}	$&$	3597	$&$	124.21	$&$	8.6\times 10^{15}	$&$	2.3\times 10^{17}	$&$	5.08	$&$	4.4\times 10^{-11}	$&$	3.60\times 10^{-4}	$&$	3.0\times 10^{-9}	$	\\ 	\hline
$	2200	$&$	1,090,000	$&$	25	$&$	173.3	$&$	594	$&$	595	$&$	2401	$&$	{\bf 3103}	$&$	3469	$&$	124.82	$&$	1.1\times 10^{16}	$&$	1.9\times 10^{17}	$&$	0.26	$&$	1.2\times 10^{-10}	$&$	3.56\times 10^{-4}	$&$	3.1\times 10^{-9}	$	\\ 	\hline
$	2600	$&$	160,000	$&$	10	$&$	172.9	$&$	671	$&$	906	$&$	2817	$&$	{\bf 3511}	$&$	4213	$&$	125.19	$&$	8.7\times 10^{15}	$&$	2.2\times 10^{17}	$&$	6.90	$&$	3.2\times 10^{-11}	$&$	3.61\times 10^{-4}	$&$	3.1\times 10^{-9}	$	\\ 	\hline
$	2950	$&$	340,000	$&$	10	$&$	172.7	$&$	787	$&$	1030	$&$	3179	$&$	{\bf 4015}	$&$	4678	$&$	125.30	$&$	8.8\times 10^{15}	$&$	1.9\times 10^{17}	$&$	8.64	$&$	2.5\times 10^{-11}	$&$	3.61\times 10^{-4}	$&$	3.0\times 10^{-9}	$	\\ 	\hline
$	3700	$&$	580,000	$&$	10	$&$	173.1	$&$	1015	$&$	1293	$&$	3950	$&$	{\bf 5026}	$&$	5742	$&$	126.39	$&$	9.0\times 10^{15}	$&$	1.7\times 10^{17}	$&$	13.01	$&$	1.7\times 10^{-11}	$&$	3.61\times 10^{-4}	$&$	3.1\times 10^{-9}	$	\\ 	\hline
$	4550	$&$	250,000	$&$	9	$&$	172.5	$&$	1227	$&$	1601	$&$	4855	$&$	{\bf 6008}	$&$	7087	$&$	126.66	$&$	9.1\times 10^{15}	$&$	1.9\times 10^{17}	$&$	19.30	$&$	9.7\times 10^{-12}	$&$	3.61\times 10^{-4}	$&$	3.0\times 10^{-9}	$	\\ 	\hline
$	5350	$&$	520,000	$&$	9	$&$	172.4	$&$	1490	$&$	1883	$&$	5664	$&$	{\bf 7103}	$&$	8147	$&$	126.90	$&$	9.2\times 10^{15}	$&$	1.7\times 10^{17}	$&$	25.82	$&$	7.2\times 10^{-12}	$&$	3.60\times 10^{-4}	$&$	3.0\times 10^{-9}	$	\\ 	\hline
\end{tabular}
\end{table*}

The measured light Higgs boson mass is given by $M_h = 125.1 \pm 0.14$~ GeV~\cite{Aad:2012tfa,Chatrchyan:2012xdj,Tanabashi:2018oca}, but for our applied constraint we allow for a 2$\sigma$ experimental uncertainty over and above a theoretical uncertainty of about 1.5~GeV. The approximate combined experimental and theoretical error gives a practical but firm constraint of $124 \le M_h < 127$~ GeV. The vector-like flippons will couple to the Higgs fields via the flippon Yukawa coupling, however, the flippon contribution to $M_h$ is inversely proportional to the vector-like decoupling scale $M_V$~\cite{Huo:2011zt}. For $M_V \gtrsim 5$~TeV, the flippon contribution quickly approaches a negligible value, hence the flippon coupling to the Higgs field can be neglected in this work with no impact to the final results.

A comparison of the No-Scale \fsu5 model is undertaken against the SUSY contribution to the rare-decay processes, namely, the branching ratio of the rare b-quark decay of $Br(b \to s \gamma) = (3.43 \pm 0.21^{stat}~ ±\pm 0.24^{th} \pm 0.07^{sys}) \times 10^{-4}$~\cite{HFAG}, the branching ratio of the rare B-meson decay to a dimuon of $Br(B_s^0 \to \mu^+ \mu^-) = (2.9 \pm 0.7 \pm 0.29^{th}) \times 10^{-9}$~\cite{CMS:2014xfa}, and the 3$\sigma$ intervals around the Standard Model result and experimental measurement of the SUSY contribution to the anomalous magnetic moment of the muon of $-17.7 \times10^{-10} \le \Delta a_{\mu} \le 43.8 \times 10^{-10}$~\cite{Aoyama:2012wk}. All points, regardless of $\Omega_{\widetilde{\chi}_1^0} h^2$, easily satisfy all three rare-decay constraints.

Approximately 25 million points were scanned for this study. The SUSY mass and Higgs spectra, dark matter relic density, and rare-decay processes were computed with {\tt MicrOMEGAs~2.1}~\cite{Belanger:2008sj}, utilizing a proprietary mpi modification of the {\tt SuSpect~2.34}~\cite{Djouadi:2002ze} codebase to run No-Scale ${\cal F}$-$SU(5)$ enhanced RGEs. The Particle Data Group~\cite{Tanabashi:2018oca} world average for the strong coupling constant is $\alpha_S (M_Z) = 0.1181 \pm 0.0011$ at 1$\sigma$, and we employ $\alpha_S = 0.1184$ in our calculations.

\section{Phenomenology}

The withdrawal of the constraint on the computed LSP relic density $\Omega_{\widetilde{\chi}_1^0} h^2$ and then relying upon the master coupling $\lambda_6$ to dilute the relic density down to the correct value does indeed open up a vast new region of viable No-Scale \fsu5 parameter space. The proportional dependence of the SUSY spectrum on $M_{1/2}$ is visually confirmed in FIG.~\ref{fig:m12}. Given the lack of a restriction on $\Omega_{\widetilde{\chi}_1^0} h^2$, the linear relationship is not quite as sharp as prior studies~\cite{Li:2016bww}, but it nonetheless prevails and remains rather strong. Evidently, the gluino mass in FIG.~\ref{fig:m12} could continue uninterrupted indefinitely, provided it may accommodate a world average top quark mass. This is not the case though, as FIG.~\ref{fig:mt} unequivocally illuminates. As explained, the largeness of the SUSY spectrum, and therefore the light stop, ensures a decreasing top quark mass in order to preside within our Higgs mass constraint. The spectra cannot support a top mass within its own world average limits~\cite{Tanabashi:2018oca} once the gluino reaches about $M_{\widetilde{g}} \simeq 7.5$~TeV and the LSP is $M_{\widetilde{\chi}_1^0} \simeq 1.6$~TeV, as FIG.~\ref{fig:mt} displays. Note that the sparseness of points at the upper left and lower right of FIG.~\ref{fig:mt} arises from limiting the calculations to $1200 \le M_{1/2} \le 5500$~GeV for pragmatic reasons. Boosting the upper $M_{1/2}$ limit would surely increase the density of points at large $M_{1/2}$ and small $M_t$, though the trend and conclusion is unmistakable nevertheless.

A dozen benchmarks points are presented in TABLE~\ref{tab:spectra}, representative of the full region of viable model space. The benchmarks points are more concentrated within the LHC's near future accessibility of about $M_{\widetilde{g}} \lesssim 3$~TeV, though characteristics of the far heavier region are submitted as well. Interestingly, the maximum tan$\beta$ for any viable point is 25, with the massive gluino region $M_{\widetilde{g}} \gtrsim 3$~TeV necessarily requiring tan$\beta \sim 10$.

The obvious demarcation point at $M_{\widetilde{g}} \simeq 7.5$~TeV illustrated in FIG.~\ref{fig:mt} does usefully serve as an approximate upper boundary on the gluino mass in the No-Scale \fsu5 model. Of course, this mass scale is primarily provided for reference since it is far beyond the reach of the LHC Run 2 and certainly the LHC in any configuration of future upgrades. Pair production of such massive gluinos most likely will require at least a 100~TeV particle collider. In the disconcerting event no SUSY discovery is celebrated at the LHC in the forthcoming years, the deeply natural attributes of No-Scale \fsu5 can serve as one component of justification for a powerful 100~TeV collider in the coming decades.

%%%%%%%%%%%%%%%%%%%%%%%%%%%%%%%%%%%%%%%%%%%%%%%%%%%%%%%%%%%%%%%%%%%%%%%%%%%%

\section{Acknowledgments}

Portions of this research were conducted with high performance computational resources provided 
by the Louisiana Optical Network Infrastructure (http://www.loni.org). This research was supported 
in part by the Projects 11475238, 11647601, and 11875062 supported 
by the National Natural Science Foundation of China (TL), 
by the Key Research Program of Frontier Science, Chinese Academy of Sciences (TL),
and by the DOE grant DE-FG02-13ER42020 (DVN). 

%%%%%%%%%%%%%%%%%%%%%%%%%%%%%%%%%%%%%%%%%%%%%%%%%%%%%%%%%%%%%%%%%%%%%%%%%%%%

\bibliography{bibliography}

\end{document}